\documentstyle[pra,aps,amsfonts,twocolumn]{revtex}
\newcommand{\trace}{\mathop{\rm Tr}\nolimits}
\newcommand{\bra}[1]{\langle#1|}
\newcommand{\ket}[1]{|#1\rangle}

\newcommand{\uparr}{\uparrow}
\newcommand{\dnarr}{\downarrow}
\newcommand{\qed}{\hfill$\Box$}
\newcommand{\C}{{\Bbb C}}
\newcommand{\identity}{\openone}

\newtheorem{theorem}{Theorem}
\newtheorem{lemma}{Lemma}

\begin{document}
\draft
\title{Variational Characterisations of Separability and 
Entanglement of Formation}
\author{Koenraad Audenaert\cite{KAmail},
Frank Verstraete\cite{FVmail},
Bart De Moor\cite{BDMmail}}
\address{Katholieke Universiteit Leuven, \\
Dept. of Electrical Engineering (ESAT) \\
Research Group SISTA \\
Kard. Mercierlaan 94 \\
B-3001 Leuven, Belgium
}
\date{\today}
\maketitle
\begin{abstract}
In this paper we develop a mathematical framework for the characterisation of 
separability and entanglement of formation (EoF) of general bipartite states. 
These characterisations are of the variational kind, meaning that separability 
and EoF are given in terms of a function which is to be minimized over the 
manifold of unitary matrices. A major benefit of such a characterisation
is that it directly leads to a numerical procedure for calculating EoF.
We present an efficient minimisation algorithm and an apply it
to the bound entangled $3\times3$ Horodecki states;
we show that their EoF is very low and that their distance to the set of 
separable states is also very low. Within the same variational framework we 
rephrase the results by Wootters (W. Wootters, Phys.~Rev.~Lett. {\bf 80}, 
2245 (1998)) on EoF for $2\times2$ states and present progress in generalising 
these results to higher dimensional systems.
\end{abstract}
\section{Introduction}
A problem which has received considerable attention in the last few years is 
to find necessary and sufficient conditions under which a quantum state of a 
composite system is separable. The example extraordinaire of a non-separable
state is a pair of 2-level particles in a singlet state, a so-called EPR-pair, 
named after Einstein, Podolsky and Rosen, who used this sort of state to show 
that quantum mechanics exhibits strong non-local correlations which seem to 
violate the relativity principle. 

A separable state of a composite system can be written 
as the direct product of the subsystem states: 
$\ket{\Psi_{AB}} = \ket{\Psi_A}\otimes\ket{\Psi_B}$. 
A non-separable state, or {\em entangled} state cannot be decomposed in this 
way; e.g., the singlet state 
$(\ket{\uparr}\ket{\dnarr} - \ket{\dnarr}\ket{\uparr})/\sqrt{2}$ consists of a 
superposition of separable states but is itself not separable. 

Nowadays, the 
importance of entangled states goes beyond a mere fundamental interest, since 
EPR-pairs are the basic resources of quantum techniques such as quantum 
cryptography, quantum teleportation and quantum error correction. A mixed 
state is separable iff its density matrix can be written as a convex linear 
combination of pure product states; for a bipartite system this reads:
\begin{equation} 
\label{eqn1}
\rho = \sum_{k=1}^K w_k \ket{u^k}\bra{u^k} \otimes \ket{v^k}\bra{v^k}.
\end{equation}
The separability problem consists of finding a criterion for checking whether 
such a decomposition is possible for a given state. 

Despite the simple formulation of this 
problem, a complete solution has to this date not been found. An important 
achievement was the discovery by Peres of a necessary condition for 
separability \cite{peres96}. He noted that the {\em partial transposition} of 
a separable state still has non-negative eigenvalues, just as the original 
state. Thus, if the partial transposition of state $\rho$ is not a state 
(i.e.\ does not have non-negative eigenvalues summing to one),then $\rho$ is 
not separable (i.e.\ is an entangled state). The importance of this criterion 
was soon realised when M., P.\ and R.\ Horodecki proved \cite{horodecki96a} 
that it is also a sufficient criterion for $2\times2$ and $2\times3$ systems.
For an introduction to recent results on this subject, see, e.g., 
\cite{sanpera00}.

If a state is entangled, one could ask for a measure of the amount of 
entanglement. For pure states, a measure generally agreed upon is the von
Neumann subsystem entropy: the entropy of the partial trace of the state
projector. For mixed states, the situation is much more difficult. Not only
is there no single measure of entanglement which is suited for every purpose,
but calculating the values of the different proposed measures and proving
statements about them is exceedingly difficult.
Among the proposed measures are the entanglement of formation \cite{bennett96}, 
the entanglement of distillation \cite{bennett96} and
relative entropy of entanglement \cite{vedral97}.

In this paper, we focus on separability, on entanglement of formation (EoF)
and on the related concept of concurrence.
All these subjects are related, because states are separable if and only if
their EoF is zero. A closed form expression exists for the EoF of $2\times2$
systems in terms of their concurrence \cite{wootters97}. A closed form
expression also exists for isotropic states of general systems \cite{terhal00}.

The purpose of this paper is to give variational characterisations of
separability and EoF for general (i.e.\ any dimensions) bipartite states.
Such a characterisation is of the form $Q(\rho)=\min_T f(\rho,T)$, that is: 
the state property under study can be found as the minimal value of some 
specific function over the manifold of unitary matrices $T$. In section II it
will be shown how this can be done. The language of section II is matrix 
analysis, not only because this allows to state the results in a most succinct 
way but also because it gives clues towards generalisations.

The greatest benefit of a variational 
characterisation is that it directly yields a method for actually calculating
the state property $Q$, albeit in a numerical fashion, using a minimisation
procedure. In section III we describe the procedure which we have used, and some
interesting results we have obtained with it.
\section{Variational characterisations}
It is well known that mixed states can be realised by an ensemble of pure 
states in an infinite number of ways.
The determination of the separability of a state and the determination of its
entanglement of formation have in common that a particular realisation of a 
state has to be found such that some property holds for all pure states in
that realisation. In order to find this optimal realisation, it is of 
considerable interest to have a mathematically elegant way of ``generating''
all possible realisations of a state. In section \ref{ss21} we will recollect
a result by Hughston, Jozsa and Wootters that any realisation of a state 
is related to the eigenvalue decomposition of the state via some 
right-unitary matrix.

The required property for separability is that all pure states in the 
realisation must be product states. In section \ref{ss22} we give a number
of useful mathematical expressions for this property. This then leads to a 
variational characterisation of separability, the topic of section \ref{ss23}.
For calculating
the EoF of the state, the property of the optimal realisation is that the
so-called average entanglement of the realisation is minimal. This property
and an ensuing variational characterisation of EoF will be discussed in 
section \ref{ss24}.

In this way, searching all possible realisations for some
property amounts to passing through all right-unitary matrices and test the
property in question.
However, this would be a very impractical way to determine separability or EoF
if there would not be some bound on the dimension of these matrices, or, which
is the same thing, on the number of pure states in the optimal realisation. 
Luckily, such a bound exists.
In the case of separability, Horodecki proved \cite{horodecki97} that 
$(N_1N_2)^2$ pure states (or less) suffice, where $N_1$ and $N_2$ are the 
dimensions of the subsystem Hilbert spaces. Uhlmann \cite{uhlmann} proved 
that a similar bound holds for the determination of EoF: the number of pure
states in the optimal realisation need not be larger than the square of the 
rank of the state.

In section \ref{ss26} we discuss the so-called concurrence of a state,
a quantity which is closely related to the EoF. We give an alternative proof of
an important result on the concurrence of $2\times2$ states by Wootters 
\cite{wootters97}. One of the virtues of this alternative proof is that it
yields an additional result on the exact amount of pure states in the optimal
realisation. We then report some progress in generalising the concurrence
concept to higher-dimensional bipartite states.

In appendix A, finally, a method is described for reducing the set of 
unitary matrices which has to be examined, in the case of separability
testing. Under some circumstances, this method directly yields an optimal 
realisation, without any need for searching. We have not yet investigated 
whether this method is applicable to the EoF case.
\subsection{Relation between different realisations of a state}
\label{ss21}
Consider a rank-$R$ state $\rho$ in an $N_1\times N_2$-dimensional Hilbert 
space, realised by an ensemble $\{w_k,\ket{\psi^k}\}_{k=1}^K$, where the $w_k$ 
are the mixing weights of the $K$ pure state vectors $\ket{\psi^k}$. The 
number $K$ is called the {\em cardinality} of the ensemble. Necessarily, $K$ 
cannot be smaller than the rank $R$. Since there generally are an infinite 
number of ensembles realising a particular mixed state, we are free to choose 
$K$ larger than $R$ if this suits our purposes. It will turn out that
sometimes we will even be forced to take $K>R$.

Thus:
$\rho = \sum_{k=1}^K w_k \ket{\psi^k}\bra{\psi^k}$, 
or $\rho = \Psi W\Psi^\dagger$, where $W$ is a $K\times K$ diagonal matrix 
with $W_{kk} = w_k$, and the columns of $\Psi$ are the $K$ vectors $\psi^k$.
This decomposition of $\rho$ is reminiscent of the eigenvalue decomposition of 
$\rho$: $\rho = \Phi M\Phi^\dagger$, where $M$ is an $R\times R$ diagonal 
matrix whose diagonal elements are the eigenvalues of $\rho$, and the columns 
of $\Phi$ are the $R$ eigenvectors. Since $\rho$ is Hermitian, $\Phi$ is a 
unitary matrix.

It can now easily be proven that these two decompositions must be related by
an $R\times K$ right-unitary matrix $T$; this has been done first by 
Hughston, Jozsa and Wootters \cite{hughston93}.
\begin{lemma}
For a general state $\rho$, with eigenvalue decomposition 
$\rho = \Phi M\Phi^\dagger$, 
there is a matrix $\Psi$ and a non-negative diagonal matrix $W$ such that 
$\rho = \Psi W\Psi^\dagger$ iff there is an $R\times K$ matrix $T$ such that:
\begin{equation} 
\label{eq:1}
\Psi W^{1/2} = \Phi M^{1/2} T, \mbox{ with } TT^\dagger=\identity_R
\end{equation}
\end{lemma}
Right-unitarity of the matrix 
$T$ means that a unitary $K\times K$ matrix $T'$ exists such that $T$ consists 
of $R$ row vectors of $T$; that is, the $R$ row vectors of $T$ form an 
orthonormal set in $\C^K$ and the $K$ column vectors are projections of an 
orthonormal basis in $\C^K$ onto an $R$-dimensional subspace.
Stated in matrix algebraic terms, the proof becomes very simple:

{\em Proof.}
First of all, it is obvious that $\Phi M\Phi^\dagger = \Psi W\Psi^\dagger$ 
follows directly from (\ref{eq:1}). Conversely, denote $X = \Psi W^{1/2}$ 
and consider the singular value decomposition of $X$: $X=U\Sigma V$,
where $U$ is a unitary $R\times R$ matrix, $V$ a right-unitary $R\times K$ 
matrix and $\Sigma$ a diagonal $R\times R$ matrix with non-negative diagonal 
elements. From $\Phi M\Phi^\dagger = \Psi W\Psi^\dagger$ we get
$\Phi M\Phi^\dagger = U \Sigma^2 U^\dagger$. Since both $M$ and $\Sigma$ are 
positive semidefinite, it follows that $\Sigma = U^\dagger \Phi M^{1/2} 
\Phi^\dagger U$ so that $X = \Phi M^{1/2} \Phi^\dagger U V$. This is precisely 
equation(\ref{eq:1}), with $T=\Phi^\dagger U V$. 
\qed

{\em Remark.}
It is noteworthy that the elements of $W$ and $M$ are related to each other
independently of $\Phi$ and $\Psi$:
$$
w_k = (T^\dagger M T)_{kk}.
$$
This follows from the observations that $\Phi$ is unitary and that the columns
of $\Psi$ have norm one.
\subsection{Characterisation of product states}
\label{ss22}
A state of an $N_1\times N_2$ system is separable iff there exists a realising 
ensemble consisting solely of product vectors $\psi=\psi^1\otimes\psi^2$, with 
$\psi^1\in{\cal H}_1$ and $\psi^2\in{\cal H}_2$ (in this paper we use 
superscripts for enumerating vectors, and subscripts for denoting vector 
components). 
Product vectors can be characterised easily by rearranging their components 
in matrix form. For an $N_1 N_2$-vector $x$, let $\tilde{x}$ be an 
$N_1\times N_2$ matrix such that $x = \sum_{i,j}\tilde{x}_{ij}e^i\otimes e^j.$
For product vectors this gives:
$$
\psi^k = \alpha^k \otimes \beta^k \longrightarrow \tilde{\psi}^k 
= \alpha^k (\beta^k)^T.
$$
Obviously, product vectors are characterised by the condition that the rank of 
$\tilde{\psi}$ is 1. A necessary and sufficient condition for this is that all 
$2\times 2$ minors of $\tilde{\psi}$ must be zero, or, more succinctly, that 
the {\em second compound matrix} of $\tilde{\psi}$ must be zero: 
$C_2(\tilde{\psi}) = 0$ \cite{horn85}. The second compound matrix 
of an $N_1\times N_2$ matrix is an $(N_1(N_1-1)/2) \times (N_2(N_2-1)/2)$ 
matrix with elements:
$$
C_2(A)_{(ii'),(jj')} = A_{ij} A_{i'j'} - A_{ij'} A_{i'j}, \mbox{ } i<i', j<j'.
$$
The elements of $C_2$ are all possible $2\times 2$ minors of $A$. 
The second compound matrix 
has a lot of useful properties,such as: $C_2(AB)=C_2(A)C_2(B)$, 
$C_2(\identity_n)=\identity_{n(n-1)/2}$ and $C_2(A^{-1}) = (C_2(A))^{-1}$ 
\cite{horn85}. 

For practical applications it is sometimes better to consider a $(N_1-1)\times
(N_2-1)$ submatrix of $C_2$, the one containing the elements 
$C_2(A)_{(i,i+1),(j,j+1)}$ only. 
It is easily seen that the vanishing of this submatrix is already sufficient 
for $A$ being of rank 1. 

From the expression for the second compound matrix, which is quadratic in $A$,
it will prove useful to construct a bilinear function of two $N_1\times N_2$ 
matrices, denoted ${\cal C}(A,B)$:
$$
{\cal C}(A,B)_{(ii'),(jj')} = A_{ij} B_{i'j'} - A_{ij'} B_{i'j}, 
\mbox{ } i<i', j<j'.
$$
Obviously, ${\cal C}(A,A)=C_2(A)$, so that ${\cal C}(A,A)=0$ if and only if 
$A$ has rank 1. More specifically, we can apply this to the state vectors 
$\psi^k$: $\psi^k$ is a product vector iff ${\cal C}(\tilde{\psi}^k,
\tilde{\psi}^k)=0$. 

In the following, we will only use a symmetrised version of ${\cal C}$, which 
we will denote by
$$
C(\psi^k,\psi^l) = {\cal C}(\tilde{\psi}^k,\tilde{\psi}^l) + 
{\cal C}(\tilde{\psi}^l,\tilde{\psi}^k).
$$
Since this is a bilinear function in the elements of $\Psi$, we can express 
this in matrix notation:
$$
C(\psi^k,\psi^l)_{(\alpha)} = (\Psi^T S^{(\alpha)}\Psi)_{kl},
$$
where the notation $(\alpha)$ is a shorthand for the index tuple $(i,i',j,j')$.
The matrices $S^{(\alpha)}$, which we call {\em indicator matrices}, are 
defined as
\begin{eqnarray*}
S^{(\alpha)}_{(ij),(i'j')} &=& S^{(\alpha)}_{(i'j'),(ij)} = 1 \\
S^{(\alpha)}_{(ij'),(i'j)} &=& S^{(\alpha)}_{(i'j),(ij')} = -1
\end{eqnarray*}
all other elements being zero. Note that all $S$ have rank equal to 4. 
For the case of $2\times2$-systems, there is only 
one indicator matrix; it is equal to $\sigma_y\otimes\sigma_y$, corresponding 
to a spin-flip operator \cite{wootters97}.
\subsection{Condition for separability}
\label{ss23}
We can now formulate a general necessary and sufficient condition for the 
separability of a mixed state. As mentioned before, the state 
$\rho = \Phi M\Phi^\dagger$ is separable iff there exists a decomposition 
$\rho = \Psi W\Psi^\dagger$, with $\Psi W^{1/2} = \Phi M^{1/2} T$, such that 
all $\psi^k$ are product states, or $C(\psi^k,\psi^l)=0$, for all $k=l$.

Now:
\begin{eqnarray}
C(\psi^k,\psi^l) &=& C(\sqrt{w_k}\psi^k,\sqrt{w_l}\psi^l)/\sqrt{w_k w_l} 
\nonumber \\
& = & \sum_{p,q=1}^R \frac{T_{pk} T_{ql}}{\sqrt{w_k w_l}} 
      C(\sqrt{m_p}\phi^p,\sqrt{m_q}\phi^q), 
\label{eq:4}
\end{eqnarray}
where we have used bilinearity of the form $C$. Given the eigenvalue 
decomposition
of $\rho$, the entity $C(\sqrt{m_p}\phi^p,\sqrt{m_q}\phi^q)$ can be calculated 
straightforwardly. Let us organise its components into a set of matrices 
$A^{(\alpha)}$:
\begin{equation} 
\label{eq:amatrix}
A^{(\alpha)}_{pq} = C(\sqrt{m_p}\phi^p,\sqrt{m_q}\phi^q)_\alpha 
= \sqrt{M} \Phi^T S^{(\alpha)} \Phi\sqrt{M}.
\end{equation}
Using this notation, (\ref{eq:4}) can be written concisely as 
$$
C(\psi^k,\psi^l) = (T^T A^{(\alpha)} T)_{lk}/ \sqrt{w_k w_l}.
$$
The state is therefore separable iff we can find an $R\times K$ matrix $T$, 
with $K\ge R$, such that
\begin{equation} 
\label{eq:sep}
\left\{
\begin{array}{l}
	TT^\dagger=\identity_R \\
	C_2(\tilde{\psi}^k) = (T^T A^{(\alpha)} T)_{kk} = 0,
 \mbox{ } \forall \alpha, k.
\end{array}
\right.
\end{equation}
Here, $k$ ranges from 1 to $K$, and $\alpha$ enumerates all tuples of indices 
$(i,i',j,j')$ with $1\le i<i'\le N_1$ and $1\le j<j'\le N_2$. As noted before,
it is also sufficient to consider only the tuples $(i,i+1,j,j+1)$.

Testing separability requires that the system (\ref{eq:sep}) be solved for $T$.
Another approach, however, is to consider $(T^T A^{(\alpha)} T)_{kk}$ as 
entries
of a matrix indexed by $\alpha$ and $k$ and to try to minimise a matrix norm of
this matrix. The state is then separable iff this minimum is zero. One can use 
whatever matrix norm one prefers, e.g.\ the Hilbert-Schmidt norm (also called 
Frobenius norm or $l_2$-norm) 
$||A||_2^2 = \sum_{i,j} |A_{i,j}|^2 = \trace A A^\dagger$. Thus $\rho$ is 
separable iff
\begin{equation}
\label{eq:sepmin}
\min_{T,K} \sum_{\alpha,k} |(T^T A^{(\alpha)} T)_{kk}|^2 = 0,
\end{equation}
where the minimum has to be taken over all $K\ge R$ and all $R\times K$ 
matrices $T$ for which $TT^\dagger=\identity_R$. The minimal $K$ is called the 
{\em cardinality} of the state. 

One can also use the $l_1$ norm (sum of absolute values) and minimise
$\sum_{\alpha,k} |(T^T A^{(\alpha)} T)_{kk}|$. For $2\times 2$ systems the 
$l_1$ norm is the {\em average concurrence} of the ensemble, as introduced 
by Wootters in \cite{wootters97}, and the minimum is the {\em concurrence} of 
the state $\rho$.
Note that in the context of separability testing it does not matter whether 
one uses $(T^T A^{(\alpha)} T)_{kk}$ or $(T^T A^{(\alpha)} T)_{kk}/w_k$.

To end this paragraph, we derive an alternative expression for the $l_2$ 
norm $||(C_2(\tilde{\psi}^k))_k||_2$.
Define $B^k=\tilde{\psi}^k (\tilde{\psi}^k)^\dagger$, with eigenvalue 
decomposition 
$B^k = U^k\Sigma^k U^{k\dagger}$ (with $\Sigma^k = \mbox{Diag}(\sigma^k_i)$).
Using the properties of $C_2$ we find
\begin{eqnarray*}
||(C_2(\tilde{\psi}^k))_{k=1}^n||_2^2 
&=& \sum_k \trace(C_2(\tilde{\psi}^k) C_2(\tilde{\psi}^k)^\dagger) \\
&=& \sum_k \trace C_2(B^k) = \sum_k \trace C_2(\Sigma^k) \\
&=& \sum_k \sum_{i<j} \sigma^k_i\sigma^k_j \\
&=& \frac{1}{2}\sum_k (\sum_{i,j} \sigma^k_i\sigma^k_j - 
     \sum_i (\sigma^k_i)^2) \\
&=& \frac{1}{2}\sum_k (\sum_i \sigma^k_i)^2 - \sum_i (\sigma^k_i)^2 \\
&=& \frac{1}{2}\sum_k (\trace \Sigma^k)^2 - \trace(\Sigma^k)^2 \\
&=& \frac{1}{2}\sum_k (\trace B^k)^2 - \trace(B^k)^2.
\end{eqnarray*}
This result can be interpreted easily: a positive definite hermitian matrix is
rank 1 iff the square of its trace equals the trace of its square.
\subsection{Entanglement of formation}
\label{ss24}
Within the same framework, we can also give a variational characterisation of 
the entanglement of formation $E(\rho)$ (EoF) of a mixed state $\rho$. This 
quantity is defined as the average entanglement of the pure states in a 
realising ensemble, minimised over all possible realising ensembles
\cite{bennett96}. The von Neumann entropy $H$ of a state $\rho$ is 
$-\trace \rho \log_2 \rho$; 
introducing the function $h(x) = -x\log_2 x$, we can express $H$ as a 
function of the eigenvalues $\lambda_k$ of $\rho$: 
$H(\rho) = \sum_k h(\lambda_k)$.
The entanglement of a pure state $\psi$ of a bipartite system $(A,B)$ is the 
entropy of the partial trace of the projector of $\ket{\psi}$: 
$E(\psi) = H(\rho_A)$, with $\rho_A = \trace_B(\ket{\psi}\bra{\psi})$. 
The average entanglement of an ensemble $\{w_k,\psi^k\}$ is
$\sum_k w_k E(\psi^k)$; 
the EoF is then found as the minimal value over all ensembles realising $\rho$.

In this paragraph, we will derive an expression for $E(\rho)$ which is better 
suited for calculation. Let $\{w_k,\psi^k\}$ be the realising ensemble with 
least average entanglement and $\{m_p,\phi_p\}$ the realising ensemble
corresponding 
to the eigenvalue decomposition of $\rho$. We first express the partial trace 
of the projector of $\psi^k$ in terms of $\tilde{\psi}^k$:
$\psi^k = \sum_{i,j}\tilde{\psi}^k_{ij}e^i\otimes e^j$, hence
$\ket{\psi^k}\bra{\psi^k} 
= \sum_{i,j,p,q}\tilde{\psi}^k_{ij} (\tilde{\psi}^k_{pq})^* 
(e^i\otimes e^j)(e^p\otimes e^q)^\dagger$, and the partial trace equals
\begin{eqnarray*}
\rho_A^k &=& \trace_B(\ket{\psi^k}\bra{\psi^k}) \\
&=& \sum_{i,p}\left( \sum_q \tilde{\psi}^k_{iq} 
    (\tilde{\psi}^k_{pq})^* \right) (e^i)(e^p)^\dagger \\
&=& \tilde{\psi}^k (\tilde{\psi}^k)^\dagger.
\end{eqnarray*}
This is precisely the matrix $B^k$ from the previous paragraph.

{\em Remark:} The entropy of this partial trace matrix $\rho_A^k$ can be 
expressed in terms of the singular values of $\tilde{\psi}^k$.
Let $\tilde{\psi}^k = U^k \Sigma^k V^k$ be the singular value decomposition 
of $\tilde{\psi}^k$ (that is, the Schmidt decomposition of $\psi^k$), with 
$U^k$ unitary and $V^k$ right-unitary (supposing that $N_1\le N_2$) and 
$\Sigma^k$ a positive semidefinite diagonal matrix, then 
$\rho_A^k = U^k (\Sigma^k)^2 (U^k)^\dagger$ and 
$H(\rho_A^k) = H((\Sigma^k)^2) = -2 \sum_i (\sigma^k_i)^2 \log_2(\sigma^k_i).$

In the present framework only the eigenvectors $\phi^p$ are known, and the 
vectors $\psi^k$ are to be sought by looking for an appropriate $T$-matrix. 
We therefore want to express $\rho_A^k$ in terms of $T$ and the $\phi^p$.
We get:
\begin{eqnarray*}
w_k \rho_A^k &=& \sqrt{w_k}\tilde{\psi}^k \sqrt{w_k}(\tilde{\psi}^k)^\dagger \\
&=& \sum_{p,q=1}^R T_{pk} T_{qk}^* 
    \sqrt{m_p m_q} \tilde{\phi}^p (\tilde{\phi}^q)^\dagger.
\end{eqnarray*}
Let us use the symbol $\Delta_k(T)$ as a shorthand for the right-hand side 
of the previous expression:
$$
\Delta_k(T) = \sum_{p,q=1}^R T_{pk} T_{qk}^* \sqrt{m_p m_q} 
              \tilde{\phi}^p (\tilde{\phi}^q)^\dagger
$$
$$
\rho_A^k = \Delta_k(T)/w_k
$$
$$
w_k = \trace\Delta_k(T).
$$
The last equation follows from the fact that $\rho_A^k$ is normalised.

The EoF is thus:
\begin{eqnarray} 
E(\rho) &=& \min_{T, K} \sum_{k=1}^K w_k H(\rho_A^k) \nonumber \\
&=& \min_{T, K} \sum_{k=1}^K G(\Delta_k(T)), \label{eq:eof}
\end{eqnarray}
with
\begin{eqnarray}
G(A) &=& -\trace (A \log_2(A/\trace(A))) \label{eq:G} \\
     &=& H(A) - h(\trace(A)). \nonumber
\end{eqnarray}
The minimum has to be taken over all $K\ge R$ and all $R\times K$ matrices 
$T$ for which $TT^\dagger=\identity_R$. Note that, since a state is separable 
iff its entropy of formation is zero, equation (\ref{eq:eof}) gives an 
alternative for equation (\ref{eq:sepmin}) for testing separability.

Equation (\ref{eq:eof}) can be brought in a more suitable form if we enlarge 
the set of matrices $\tilde{\phi}^p$ with zero matrices for $p>R$. Then we 
can always use square, and therefore unitary $T$ matrices.
Following a result by Uhlmann \cite{uhlmann}, the cardinality $K$ must lie
between the rank $R$ and the square of the rank. This guarantees that the
EoF can be found by restricting oneself to finite sized $T$ matrices.
\subsection{Concurrence}
\label{ss26}
The first analytic formula for calculating EoF has been found by Wootters
\cite{wootters97} and is valid for $2\times2$ systems. A basic property
used in deriving the formula is the so-called concurrence of a state.
The concurrence is also useful for testing separability, because a $2\times2$ 
state is separable iff its concurrence equals zero. In this section we do two 
things: first we rederive Wootters' results in a shorter way, based on the
concepts we have introduced above and using an interesting theorem from 
matrix analysis. This rederivation gives hints toward the generalisation of the
concurrence concept to higher-dimensional systems, which is the second topic 
of this section.
\subsubsection{The $2\times 2$ case}
In this paragraph we give a shorter proof of Wootters' results on the EoF of 
$2\times 2$ systems \cite{wootters97}. For the case of $2\times2$ systems, 
formula (\ref{eq:sep}) becomes particularly simple, since there is only one 
$2\times 2$ minor to consider, so that there is just a single symmetric matrix 
$A^{(\alpha)}$. 

The concurrence of a pure state $\psi$ equals $C(\psi) = |\psi^T S \psi|$.
For the pure states $\psi^k$ in a decomposition of $\rho$, we get
$C(\psi^k) = |(\Psi^T S \Psi)_{kk}| = |(W^{-1/2} T^T A T W^{1/2})_{kk}|
= |(T^T A T)_{kk}|/w_k$.

The average concurrence of a realisation of $\rho$ is thus given by 
$\sum_k |(T^T A T)_{kk}|$ and the concurrence of $\rho$ is the minimal average 
concurrence over all possible realisations, i.e.\ over all possible 
right-unitary $T$. Since $A$ is symmetric, its singular value decomposition 
assumes a special form, known as the Takagi eigenvalue decomposition 
\cite{horn85}: 
$A = U^T\Sigma U$ (again, $U$ is unitary and $\Sigma$ positive 
semidefinite diagonal). 
Since we consider all possible $T$, the matrix $U$ can be absorbed in $T$, 
so that the expression for the concurrence becomes 
$\min_T \sum_k |(T^T \Sigma T)_{kk}|$. 
So, $T^T \Sigma T$ runs through all possible complex symmetric $K\times K$ 
matrices with $R$ prescribed singular values $\Sigma$ (if $K>R$ then 
$K-R$ zero singular values have to be added to $\Sigma$) and the average 
concurrence equals the sum of the moduli of the diagonal elements.

The following theorem by Thompson gives a precise relationship between the 
moduli of the diagonal elements of a complex square symmetric matrix and
its singular values \cite{thompson79} (stated without proof):
\begin{theorem}[Thompson] 
\label{th:thompson}
Let $d_1,\ldots,d_n$ be complex numbers and $s_1,\ldots,s_n$ nonnegative 
real numbers, enumerated so that $|d_1|\ge\cdots\ge|d_n|$ and 
$s_1\ge\cdots\ge s_n$. 
A complex symmetric matrix exists with $d_1,\ldots,d_n$ as its diagonal 
elements and $s_1,\ldots,s_n$ as its singular values, if and only if
$$
\sum_{i=1}^k |d_i| \le \sum_{i=1}^k s_i, \mbox{ }1\le k\le n
$$
$$
\sum_{i=1}^{k-1} |d_i| - \sum_{i=k}^n |d_i| \le 
(\sum_{i=1,i\neq k}^n s_i) - s_k, 
\mbox{ }1\le k\le n
$$
$$
\sum_{i=1}^{n-3} |d_i| -|d_{n-2}| -|d_{n-1}| -|d_n| \le 
(\sum_{i=1}^{n-2} s_i) -s_{n-1}-s_n.
$$
The last inequality does not apply when $n<3$.
\end{theorem}
The second inequality gives, for $k=1$:
$$
\sum_{i=1}^n |d_i| \ge s_1-(\sum_{i=2}^n s_i).
$$
Applied to the problem at hand, we find that the minimal average concurrence 
must be $\sigma_1 - (\sum_{i=2}^K \sigma_i)$, or zero if this quantity is 
negative. Here we have put $K=4$. Letting $K$ be larger than $4$ can give 
no improvement, since this amounts to just adding $K-4$ zero singular values,
and this does not influence the inequalities of the theorem.

If $R<4$, we could try to put $K=3$, but then the third inequality comes
into play:
$$
\sum_{i=1}^3 |d_i| \ge -(\sigma_1-(\sum_{i=2}^3 \sigma_i)),
$$
so that
$$
C(\rho)_{K=3} = |\sigma_1 - \sigma_2 - \sigma_3|.
$$
Therefore, if $R=3$ and $\sigma_1 - \sigma_2 - \sigma_3<0$, putting $K=4$ gives 
zero EoF, while $K=3$ gives non-zero EoF.
In other words, these states are separable in (at least) four product states 
($K=4$).
Furthermore, a rank 3 state is separable in three product states ($K=3$) iff
$\sigma_1 - \sigma_2 - \sigma_3=0$.

If $R=2$, we can safely put $K=2$, since then the third inequality does not 
apply.  

We have thus proven:
\begin{theorem}
The concurrence of a $2\times2$ state equals:
$$
C(\rho) = \max(0,\sigma_1 - (\sum_{i=2}^R \sigma_i)).
$$
where $\sigma_i$ are the singular values of the corresponding $A$-matrix, 
in descending order.
The optimal cardinality $K$ equals the rank $R$, except in the case when 
$R=3$ and $\sigma_1 < \sigma_2 + \sigma_3$, where the optimal $K$ is 4.
\end{theorem}
Because of the statement about the optimal cardinality, this theorem is an 
improvement over Wootter's theorem.
\subsubsection{Relation between concurrence and entanglement of formation}
For the sake of completeness, we rephrase the rest of Wootters' results
of \cite{wootters97} in the present setting.

The entanglement of a pure state is a convex, 
monotonous function ${\cal E}$ of the concurrence of the state: 
$E(\psi) = {\cal E}(C(\psi))$. Hence, the EoF, which is the average 
pure state entanglement, equals 
$$
E(\rho) = \min_T \sum_k w_k {\cal E}(|(T^T A T)_{kk}|/w_k).
$$
Because of the convexity of ${\cal E}$, this gives
$E(\rho) \ge \min_T {\cal E}(\sum_k |(T^T A T)_{kk}|)$, where equality holds
only if all quantities $|(T^T A T)_{kk}|/w_k$ are equal. Using Thompson's
theorem again and the monotonicity of ${\cal E}$, this minimum is equal to 
${\cal E}(\sigma_1 - \sum_{j>1} \sigma_j) = {\cal E}(C(\rho))$.

We therefore look for an optimal $T$ matrix, yielding minimal average 
concurrence ($C(\rho)$), and for which, additionally, all the quantities 
$|(T^T A T)_{kk}|/w_k$ are equal (and thus equal to $C(\rho)$).
There exists a $T'$ for which $\sum_k (T'^T A T')_{kk}$ is equal to $C(\rho)$;
indeed, with $A=U^T \Sigma U$, set $UT'=\mbox{Diag}(1,i,i,\ldots,i)$,
then $T'^T A T' = \mbox{Diag}(1,-1,-1,\ldots,-1)\Sigma$, and the trace of this 
matrix is $\sigma_1-(\sigma_2+\cdots+\sigma_K)$. If this quantity is positive, 
it is equal to $C(\rho)$; if not, $\rho$ is separable and we immediately have 
that all $|(T'^T A T')_{kk}|/w_k$ are equal (zero).

Concerning the non-separable states: for any orthogonal matrix $O$, 
$\trace (T'O)^T A (T'O) = \trace T'^T A T'$. As described in \cite{wootters97},
using a suitable $O$ we can make all $((T'O)^T A T'O)_{kk}$ equal to a constant
$\alpha$ times $w_k$ (exploiting the fact that $T'^T A T'$ is a real diagonal 
matrix here). Summing over $k$ then yields 
$C(\rho) = |\sum_k ((T'O)^T A T'O)_{kk} | = |\alpha \sum_k w_k| = |\alpha|$, 
so that $((T'O)^T A T'O)_{kk} = C(\rho) w_k$. Then, 
$\sum_k |((T'O)^T A T'O)_{kk}| = C(\rho)$, so that $T=T'O$ is the matrix
we were looking for. 
\subsubsection{Generalised concurrence}
According to equation (\ref{eq:sep}), a state is separable iff a right-unitary 
$T$ can be found such that the diagonal elements of every $T^T A^{(\alpha)} T$ 
are zero. In analogy with defining the average concurrence of a realisation of 
a $2\times2$ state as the $l_1$-norm of the diagonal elements of $T^T A T$, 
in the general case we can define a {\em concurrence vector} as the vector of 
$l_1$-norms of the diagonal elements of $T^T A^{(\alpha)} T$:
\begin{equation} 
\label{eq:vecconc}
C_{(\alpha)}(T) = \sum_k|(T^T A^{(\alpha)} T)_{kk}|.
\end{equation}
A state is therefore separable iff a $T$ exist such that the concurrence vector
is zero. From the previous paragraph,a necessary condition follows immediately:
\begin{equation} 
\label{eq:vecconcnec}
\sigma_1^{(\alpha)}  \le \sum_{i=2}^R \sigma_i^{(\alpha)}, 
\mbox{ } \forall (\alpha),
\end{equation}
where the $\sigma_i^{(\alpha)}$ are the singular values of $A^{(\alpha)}$, 
in descending order. 

Unfortunately, this condition is not a sufficient one for separability. 
Numerical experiments showed that criterion (\ref{eq:vecconcnec}) is 
weaker than the Peres criterion, which is a non-sufficient criterion itself. 
The main reason for this failure is that 
all the components of the vector concurrence (\ref{eq:vecconc}) must be made 
zero by one and the same $T$. Typically, however, the matrices $A^{(\alpha)}$ 
all have different singular vectors (the rows of the $U$ matrix), so that the 
$U^{(\alpha)}$ matrices in the decomposition 
$A^{(\alpha)} = U^{(\alpha)T}\Sigma^{(\alpha)}U^{(\alpha)}$ 
cannot all be absorbed in $T$ at the same time.

It is easy, however, to find a stronger criterion than criterion 
(\ref{eq:vecconcnec}): as equation (\ref{eq:vecconc}) is linear in the matrices
$A^{(\alpha)}$, the condition (\ref{eq:vecconcnec}) must hold also for every 
linear combination of the matrices $A^{(\alpha)}$. Denoting the $j$-th singular
value (descending order) of the linear combination 
$\sum_{(\alpha)} x_{(\alpha)} A^{(\alpha)}$ by $\sigma_j(x)$, it follows that 
another, and potentially stronger, necessary condition for separability is 
given by:
\begin{equation} 
\label{eq:allconc}
\max_{x\in C^M} \frac{\sigma_1(x)}{\sum_{j=2}^R\sigma_j(x)} \le 1,
\end{equation}
where $M$ is the number of tuples ${(\alpha)}$. 
Again, one could choose to consider all possible $A^{(\alpha)}$ 
or just the minimal subset with $(\alpha) = (i,i+1,j,j+1)$.

Numerical experiments now showed that criterion (\ref{eq:allconc}) is actually 
stronger than the Peres criterion, provided {\em all} $A^{(\alpha)}$ are used. 
In the next section we will give an example where condition (\ref{eq:allconc})
even seems to be sufficient for determining separability.
\section{Numerical results}
In this section we present an application of the variational characterisations
of separability and EoF. Since these characterisations involve looking for the
minimum of a function over a finite-dimensional manifold, it must be possible
to find a numerical algorithm that actually calculates that minimum.
As a result, it must be possible to calculate the EoF for {\em any} bipartite
state and, moreover, to give the optimal realisation of the state (from the
optimal $T$ matrix). In the following paragraphs, we first present in some 
detail a practical minimisation algorithm for this problem, and then apply
the algorithm to the calculation of EoF for a family of $3\times3$ states.
\subsection{Algorithm for minimisation}
Our algorithm for calculating the entanglement of formation is
based on a modified conjugate gradient minimisation procedure. Starting from an
initial point $T=T_0$, conjugate gradient algorithms iteratively seek a 
direction along which progress in minimising the objective function is optimal 
and then perform a so-called line search to actually find the minimum along 
that direction. In the present case, however, minimisation is over the unitary 
manifold. This manifold is not Euclidean, and the standard line search has to 
be replaced by a geodesic search \cite{dehaene}. A geodesic on the unitary 
manifold is a one-parameter subgroup of the unitary group:
$T(t)=T_0 \exp(t X)$, with $X$ a skew-Hermitian matrix giving the direction 
(tangent vector) of the geodesic. Through a geodesic search one looks for the 
optimal $t$ for which $g(T_0 \exp(t X))$ is minimal.

In steepest descent minimisation, the direction for the line search is 
taken to be minus the gradient of the objective function in the current point. 
Conjugate gradient methods improve on this by taking the direction of the 
previous step also in account; if not, the progress made in the previous step
could be partly undone by the new iteration. We have used a modification of the
Polak-Ribi\`ere formula for calculating the search direction \cite{fletcher}; 
the search direction for iteration $i$ is based on the gradient at the current 
point and on the search direction for the previous iteration $i-1$:
$$
X_i = -(\nabla g)_i+\gamma X_{i-1},
$$
$$
\gamma = \frac{\langle (\nabla g)_i - \tau(\nabla_g)_{i-1},(\nabla g)_i\rangle}
{\langle(\nabla g)_{i-1},(\nabla g)_{i-1}\rangle},
$$
where $\langle ,\rangle$ is the inner product of the embedding space, being in 
this case the standard Hilbert-Schmidt inner product 
$\langle x,y\rangle = \trace xy^\dagger$.
The symbol $\tau$ denotes parallel transport of the gradient vector at the 
$(i-1)$th point to the $i$th point along the geodesic \cite{dehaene}:
$$
\tau(\nabla_g)_{i-1} = 
\exp(X_{i-1}t_{i-1}/2)(\nabla_g)_{i-1} \exp(-X_{i-1}t_{i-1}/2).
$$

For the line search, we have used the method described in \cite{fletcher}, 
again modified to take into account that the search is performed along the 
geodesic $g(T_i \exp(t X_i))$.

Any minimisation algorithm actually finds local minima. To find the 
global minimum, we select a number of starting points at random and
let the minimisation algorithm work from these points. The minimum is then 
taken over all the results. While this procedure does not guarantee that the 
global minimum is actually found, we found that trying about ten starting 
points gives satisfactory results.
\subsection{Calculation of the gradient}
In this paragraph we give an analytic expression for the gradient of the target
function $g(T)$. Conjugate gradient methods perform better if an explicit
expression is given; in the absence of such an expression, the gradient has to 
be approximated numerically.

To calculate the gradient, we have to select an arbitrary direction or tangent 
vector $X$, which for the unitary manifold is a skew-Hermitian matrix. 
The geodesic on the unitary manifold along this direction and passing through 
$T_0$ is given by $T_\epsilon = T_0 \exp(\epsilon X)$, or 
$T_0(\identity+\epsilon X)$, for small $\epsilon$.
The gradient of a scalar function on the manifold can be calculated from the 
variation of the function along the geodesic, using
$$
\frac{\partial f(T_\epsilon)}{\partial \epsilon} = \langle \nabla f,X\rangle,
$$
where $\langle ,\rangle$ is the Hilbert-Schmidt inner product.

The gradient of the target function $g(T)$ is:
\begin{lemma}
$$
\left. (\nabla g(T))_{kp}\right|_{T=\identity} = 
 {\cal G}(Q^{pk},Q^{pp}) - {\cal G}(Q^{pk},Q^{kk}),
$$
where
$$
Q^{pq} = \sqrt{m_p m_q} \tilde{\phi}^p (\tilde{\phi}^q)^\dagger
$$
and 
$$
{\cal G}(B,A) = -\trace B \log_2 \frac{A}{\trace A}.
$$
\end{lemma}
The details of the calculation are given in appendix B.
\subsection{Results}
As a preliminary test, we have calculated the entanglement of formation of 
several states of a $2\times2$ system, and compared the numerical values 
with the ones obtainable from Wootters' formula. Furthermore, we considered
a one-parameter family of $3\times3$ states called isotropic states, and 
compared the numerical values with the EoF calculated from Terhal and 
Vollbrecht's formula \cite{terhal00}. In all cases, agreement was complete 
within numerical machine precision, except for some isotropic states where 
there was a very small deviation from the formula for parameter values close to
$8/9$. This can be explained by the fact that, for these parameter values, 
there are two local minima of the target function which are very close in value,
and that the minimum with lowest value has a very small ``basin of attraction''. 

The first interesting results were obtained on the Horodecki $3\times3$ states 
\cite{horodecki97}. These states were introduced to show that the Peres 
criterion is not sufficient for determining separability. These states exhibit 
{\em bound entanglement}: their entanglement of formation is non-zero, 
while their entanglement of distillation is zero (they have positive partial 
transposition). The density matrix of a Horodecki $3\times3$ state is
$$
\rho(a) = \frac{1}{1+8a}\left[
\begin{array}{ccccccccc}
a&0&0&0&a&0&0&0&a \\
0&a&0&0&0&0&0&0&0 \\
0&0&a&0&0&0&0&0&0 \\
0&0&0&a&0&0&0&0&0 \\
a&0&0&0&a&0&0&0&a \\
0&0&0&0&0&a&0&0&0 \\
0&0&0&0&0&0&b&0&c \\
0&0&0&0&0&0&0&a&0 \\
a&0&0&0&a&0&c&0&b
\end{array}\right],
$$
where $a$ is a parameter between 0 and 1, inclusively, and $b=(1+a)/2$ 
and $c=\sqrt{1-a^2}/2$. Note that, since these states are not full-rank 
(their rank is 7), and neither is their partial transpose, these states lie 
on the boundary of the set of states and also on the boundary of the set of 
bound entangled states.

The result of the calculation is shown in Fig.\ \ref{fighoro1}.
Here the entanglement of formation has been calculated of a mixture of the 
Horodecki states with the maximally mixed state: $e\rho(a)+(1-e)\identity/9$. 
In Fig.\ \ref{fighoro1}, the scale is linear, while in Fig.\ \ref{fighoro2} 
the scale is logarithmic, so that the borderline of the set of separable states
is clearly visible. The ``floor'' in the logarithmic picture at -10 is an 
artifact; the algorithm stops when the entanglement gets below $10^{-10}$.

Note from these results that the Horodecki states have a rather low 
entanglement of formation (about 0.0109 for $a=0.225$) and that their distance
to the manifold of separable states is also small ($e=0.93$ for $a=0.225$;
that is: mixing the state with just 7\% of the identity destroys all 
entanglement). At first sight, the fact that the appearance of the set of 
separable states is not convex might seem confusing. However, the parameter 
$a$ appears in a non-linear way in the density matrix so that the matrices 
lie on a non-rectilinear curve in the Euclidean state space. The figure, on 
the other hand, has $a$ as parameter and therefore gives a distorted image.

Fig.\ \ref{fighoro3} shows the entanglement of formation for the particular 
value of $a=0.225$ and for $e$ going to 1. From this figure, we are led to 
conjecture that the derivative to $e$ becomes infinite at $e=1$.

The abovementioned calculations have been performed with the cardinality $K$ 
set to 14. Fig.\ \ref{figcard} shows the effect of using different $K$ in the 
calculations; here $e=1$ and $a=0.225$. It is seen that the value $K=14$ is 
optimal for calculating the entanglement of formation in this case.

For these same Horodecki states, we have also tested the conjectured condition 
for separability (equation (\ref{eq:allconc})), based on the generalised 
concurrence. It turned out, quite surprisingly, that the condition correctly 
pinpointed all separable states, which was verified by comparing the 
results to Fig.\ \ref{fighoro2}. This leads us to hope that equation 
(\ref{eq:allconc}) might be an important step towards finding a simple and 
efficient operational criterion for testing separability.
\section{Conclusions}
We have presented a matrix analytical framework within which the questions
of separability of mixed states and calculating their entanglement of formation
can be formulated in an elegant and practical way. A main result is that, at
least in principle, it is now
be possible to calculate the EoF of any state, or determining whether it is
a separable state or not. Of course, for larger dimensions the subproblem of
minimising the respective target function becomes increasingly more time
consuming. Not only the EoF itself, but also an optimal ensemble realising the
state can be calculated.

We have extended results on the concurrence and EoF of $2\times2$ systems by 
also including the cardinality of the optimal ensembles.
More importantly, we have tried to generalise the concept of concurrence to
general systems, and have shown that this generalised concurrence has potential 
to supply a fast test for separability of general bipartite states.

In the future, we will use the presented methods to generate more numerical
results about EoF of higher-dimensional states, for example, to chart the 
``unknown territory'' of bound-entangled states, or just as a means for testing
various conjectures. Furthermore, the variational
characterisation of EoF could be useful in proving or disproving that EoF
is additive. Another interesting topic for future work is trying to prove
the conjectured sufficiency of the generalised concurrence test for 
separability.
\acknowledgements
We thank Nicolas Cerf, Henri Verschelde, Roland Puystjens, Jeroen Dehaene and 
Lieven De Lathauwer for useful discussions.
Koenraad Audenaert is postdoctoral researcher with the K.U.Leuven. 
Frank Verstraete is PhD student with the K.U.Leuven.
Bart De Moor is a Research Associate with the Flemish Fund for Scientific 
Research (FWO) and Professor Extra-ordinary at the K.U.Leuven.
This work is supported by several institutions: 
{\small
\begin{enumerate}
 \item the Flemish Government:
 	\begin{enumerate}
 	\item Research Council K.U.Leuven : 
		Concerted Research Action Mefisto-666
 	\item the FWO projects G.0240.99, G.0256.97, and Research Communities: 
		ICCoS and ANMMM 
	\item IWT projects: EUREKA 2063-IMPACT, STWW 
	\end{enumerate}
 \item the Belgian State:
 	\begin{enumerate}
 	\item IUAP P4-02 and IUAP P4-24
  	\item Sustainable Mobility Programme - Project MD/01/24
 	\end{enumerate}
 \item the European Commission:
 	\begin{enumerate}
 	\item TMR Networks: ALAPEDES and System Identification
 	\item Brite/Euram Thematic Network : NICONET
 	\end{enumerate}
 \item Industrial Contract Research : 
	ISMC, Electrabel, Laborelec, Verhaert, Europay
 \end{enumerate}
The scientific responsibility is assumed by the authors.}
\appendix
\section{Preselection of $T$ matrix}
The topic of this appendix is a method for reducing the set of $T$ matrices 
over which the minimum (\ref{eq:sep}) has to be taken in a separability test. 
In some cases the method already yields the optimal $T$ matrix without need for
performing a minimization procedure. This method is based on a method used in 
blind identification for array processing \cite{cardoso91}. 

Consider the expression 
$$
\sum_{p,q} B_{pq} A_{pq}^{(\alpha)},
$$
where $A_{pq}^{(\alpha)}$ is as defined in (\ref{eq:amatrix}), and $B_{ij}$ 
is a symmetric matrix. When we substitute equation (\ref{eq:1}) in it, we get, 
using bilinearity of $C$:
\begin{eqnarray*}
&& \sum_{p,q} B_{pq} A_{pq} \\
&=& \sum_{p,q=1}^R B_{pq} C(\sqrt{m_p} \phi^p,\sqrt{m_q} \phi^q) \\
&=& \sum_{p,q=1}^R B_{pq} \sum_{k,l=1}^K T^\dagger_{kp} T^\dagger_{lq} 
    \sqrt{w_k w_l} C(\psi^k,\psi^l) \\
&=& \sum_{k,l=1}^K \left( \sum_{p,q=1}^R B_{pq} T^\dagger_{kp} T^\dagger_{lq} 
    \right) \sqrt{w_k w_l} C(\psi^k,\psi^l) \\
&=& \sum_{k,l=1}^K (T^\dagger B T^*)_{kl} \sqrt{w_k w_l} C(\psi^k,\psi^l).
\end{eqnarray*}
Note that, just like $B$, $(T^\dagger B T^*)$ is also symmetric.

Suppose that the state $\rho$ is indeed a separable one; then there exist 
matrices $T$ leading to a product state decomposition, i.e.\ to 
$C(\psi^k,\psi^l)$ 
being identically zero for $k=l$. Consider one such $T$. There exist symmetric 
matrices $B$ for which $(T^\dagger B T^*)$ is diagonal, say equal to some 
$\Lambda$. Indeed, by right-unitarity of $T$ one just has to take 
\begin{equation} 
\label{eq:bt}
B = T \Lambda T^T.
\end{equation}
Using such a $B$ in the above expression, we find
\begin{equation} 
\label{eq:3}
\sum_{p,q} B_{pq} A_{pq}^{(\alpha)} = 0,
\end{equation}
for all $\alpha$.

We can now reverse the reasoning and say that any $T$ leading to a product 
state decomposition can be found from some symmetric $B$ that satisfies 
(\ref{eq:3}). That is, instead of searching for a $T$ in the complete set 
of unitary matrices, we only have to consider $T$ that follow, using 
(\ref{eq:bt}) and (\ref{eq:3}), from such $B$. If $T$ is square 
(that is, $K=R$), $T$ is unitary, and since $B = T \Lambda T^T$,
\begin{eqnarray*}
BB^* &=& T \Lambda T^T T^* \Lambda^* T^\dagger \\
&=& T |\Lambda|^2 T^\dagger.
\end{eqnarray*}
Hence, the column vectors of $T$ must be the eigenvectors of $BB^*$.
Given then all the symmetric matrices $B$ that satisfy equation (\ref{eq:3}), 
we only have to consider $T$ matrices whose column vectors are the eigenvectors
of one such $B$.

We will now show that under some conditions the reduced search space 
contains nothing but the optimal $T$, so that no search has to be done at all. 
In that case, one just has to take one $B$ satisfying equation (\ref{eq:3}),
and construct a $T$ from its eigenvectors. The first requirement for this 
is that the cardinality $K$ must equal the rank $R$, so that $T$ is then 
unitary; the reason is that otherwise (\ref{eq:bt}) has no unique solution.
Let us suppose that the first $P$ ($P\le K$) statevectors in the ensemble 
realising $\rho$ are product vectors: 
$\ket{\psi^k} = \ket{\alpha^k}\otimes\ket{\beta^k},1\le k\le P$.
Therefore, $C(\psi^k,\psi^k)$ will be zero for $k\le P$. Now, the matrices 
$C(\psi^k,\psi^l)$ for $k<l$ and $k=l>P$ are in general (that is: for all 
states except for a subset of measure zero) linearly independent, as long as 
the number of matrices does not exceed the number of matrix elements. If
the latter requirement is not fulfilled, then of course a dependence must
exist between the matrices. If the requirement is fulfilled then the
matrices can still be dependent provided the $K$ vectors $\psi^k$ (being
$m=KN_1N_2$ complex variables) satisfy a system of
$N_1(N_1-1)N_2(N_2-1)/4-K(K-1)/2-K+P+1$ polynomial equations of degree
$d=K(K-1)+2(K-P)$ (each equation corresponds to a minor of rank
$K(K-1)/2+K-P$ of a matrix containing $(\Psi^T S^{(\alpha)}\Psi)_{kl}$ as
elements).  Using the Schwarz-Zippel theorem \cite{zippel79}, we find that
the set of vectors obeying just one of those polynomial equations has
measure zero with respect to the set of all possible sets of $K$ vectors.
A fortiori, this also holds for the set of vectors obeying all polynomial
equations. We thus get a second requirement for the automatic optimality
of $T$, namely that the cardinality $K$ must satisfy the inequality
\begin{equation} 
\label{eq:rankcond}
\frac{K(K-1)}{2}+K-P \le \frac{N_1(N_1-1)}{2} \frac{N_2(N_2-1)}{2}.
\end{equation}
It then follows that $\sum_{p,q} B_{pq} A_{pq}$ can only be zero if 
$(T^\dagger B T^*)_{kl}=0$ for all $k\neq l$ and $k=l>P$.
In other words: $(T^\dagger B T^*)$ is necessarily a diagonal matrix 
for {\em any} $B$ satisfying (\ref{eq:3}), and any $T$ obeying (\ref{eq:bt}) 
is optimal.

We have not investigated whether this technique for reducing the search 
space is also applicable for calculating the EoF; that is, whether some 
$T$ that is optimal w.r.t.\ (\ref{eq:eof}) can be found
in the reduced search space.
\section{Calculation of the gradient of the average entanglement}
The geodesic on the unitary manifold along a direction $X$ (skew-Hermitian
matrix) and passing through $T_0$ is given by $T_\epsilon = T_0
\exp(\epsilon X)$, or $T_0(\identity+\epsilon X)$, for small $\epsilon$.
The gradient of a scalar function on the manifold can be calculated from
the variation of the function along the geodesic, using 
$$
\frac{\partial f(T_\epsilon)}{\partial \epsilon}  = \langle \nabla f,X\rangle.
$$
To avoid notational clutter, we have set $T_0$ equal to $\identity$ in the
rest of the appendix.

Let us recollect that the function of $T$ which is to be minimised is 
$g(T) = \sum_k G(\Delta_k(T))$, where
$G(A) = -\trace (A \log_2(A/\trace(A)))$ and
$\Delta_k(T) = \sum_{p,q=1}^R T_{pk} T_{qk}^* \sqrt{m_p m_q} 
    \tilde{\phi}^p (\tilde{\phi}^q)^\dagger$.

\begin{lemma}
For Hermitian $A$ and $B$,
$$
\left. \frac{\partial}{\partial\epsilon} G(A+\epsilon B)\right|_{\epsilon=0} =
 {\cal G}(B,A),
$$
where 
$$
{\cal G}(B,A) = -\trace(B\log_2 A) + \trace(B) \log_2 \trace(A).
$$
\end{lemma}
{\em Proof.} We use the following formula from \cite{horn91} (formula 6.6.31), 
which applies for a Hermitian matrix $A(t)$ function of a parameter $t$ with 
eigendecomposition $A(t)=U(t)\Lambda(t)U(t)^\dagger$, and for analytic
functions $f$: 
$$
\frac{d}{dt}f(A(t)) = U\left[ (\Delta f(\lambda_i,\lambda_j))_{ij} \circ 
U^\dagger A' U \right] U^\dagger.
$$
Here, $\circ$ is the Hadamard product and $\Delta f(\lambda_i(t),\lambda_j(t))$
are the ``divided differences''
$$
\Delta f(\lambda_i(t),\lambda_j(t)) =
\left\{\begin{array}{l} 
 \frac{f(\lambda_i(t)) - f(\lambda_j(t))}{\lambda_i(t) - \lambda_j(t)}, 
 \mbox{ for } i\neq j \\
 f'(\lambda_i(t)),
 \mbox{ for } i=j.
 \end{array}
\right.
$$
For $A(t) = A+t B$, it follows that:
\begin{eqnarray*}
\left. \frac{d}{dt}\trace f(A(t))\right|_{t=0} &=& 
\sum_i \Delta f(\lambda_i(t),\lambda_i(t)) (U^\dagger B U)_{ii} \\
&=& \trace f'(\Lambda) U^\dagger B U \\
&=& \trace f'(A) B.
\end{eqnarray*}
Setting $f(x) = h(x) = -x \log_2(x)$, so that $f(A) = H(A)$, we have 
$f'(x) = -(1+\ln x)/\ln2$ and
$$
\left. \frac{d}{dt}\trace H(A+t B)\right|_{t=0} 
= -\trace(\identity +\ln A)B/\ln 2.
$$
Furthermore,
$$
\left. \frac{d}{dt} h(A+t B)\right|_{t=0} 
= -(1 +\ln \trace A)\trace B/\ln 2,
$$
so that the lemma follows. 
\qed

Proceeding in a similar fashion, we can expand $\Delta_k(T_\epsilon)$ up to 
first order in $\epsilon$. Putting
$Q^{pq} = \sqrt{m_p m_q}\tilde{\phi}^p (\tilde{\phi}^q)^\dagger$:
\begin{eqnarray*}
\Delta_k(T_\epsilon) &=& \sum_{p,q} T_{pk} T_{qk}^* Q^{pq} \\
&=& \sum_{p,q} \left( \delta_{pk}\delta_{qk} + \epsilon (X_{pk}\delta_{qk}+
\delta_{pk}X^*_{qk}) \right) Q^{pq} \\
&=& Q^{kk} +\epsilon \sum_p (X_{pk} Q^{pk} - X_{kp} Q^{kp}),
\end{eqnarray*}
where we have used the fact that $X$ is skew-Hermitian.
Inserting this expression in 
$\left.\frac{\partial}{\partial\epsilon} \sum_k G(\Delta_k(T_\epsilon)) 
\right|_{\epsilon=0}$ we see that $Q^{kk}$ serves the role of ``$A$'' 
and $\sum_p (X_{pk} Q^{pk} - X_{kp} Q^{kp})$ that of ``$B$''.
Exploiting linearity of ${\cal G}$ with respect to its first argument, 
we arrive at the expression
$$
\frac{\partial g(T_\epsilon)}{\partial \epsilon}=
\sum_{p,k} X_{pk} ({\cal G}(Q^{pk},Q^{kk}) - {\cal G}(Q^{pk},Q^{pp}))
$$
(in the last term we have interchanged the indices $k$ and $p$).
Therefore, 
$$
\left. (\nabla g(T))_{kp}\right|_{T=\identity} = 
{\cal G}(Q^{pk},Q^{pp}) - {\cal G}(Q^{pk},Q^{kk}).
$$

\begin{figure}
\caption{Entanglement of formation for Horodecki states in function of 
$a$ and $e$; linear scale.}
\label{fighoro1}
\end{figure}
\begin{figure}
\caption{Entanglement of formation for Horodecki states in function of 
$a$ and $e$; logarithmic scale.}
\label{fighoro2}
\end{figure}
\begin{figure}
\caption{Entanglement of formation for Horodecki state $a=0.225$ in 
function of $e$; linear scale.}
\label{fighoro3}
\end{figure}
\begin{figure}
\caption{Effect of cardinality on calculation of entanglement of formation.}
\label{figcard}
\end{figure}

\begin{thebibliography}{99}
\bibitem[*]{KAmail} koen.audenaert@esat.kuleuven.ac.be
\bibitem[\dagger]{FVmail} frank.verstraete@esat.kuleuven.ac.be
\bibitem[\ddagger]{BDMmail} bart.demoor@esat.kuleuven.ac.be
%
\bibitem{peres96} A. Peres, Phys.~Rev.~Lett. {\bf 77}, 1413 (1996).
\bibitem{horodecki96a} M., P. and R. Horodecki, Phys.~Lett.~A {\bf 223}, 
1 (1996).
\bibitem{sanpera00} M. Lewenstein, D. Bru\ss, J. Cirac, B. Kraus, M. Kus, 
J. Samsonowicz, A. Sanpera and R. Tarrach, quant-ph/0006064 (2000).
\bibitem{bennett96} C. Bennett, D. DiVincenzo, J. Smolin and W. Wootters, 
Phys.~Rev.~A{\bf 54}, 3824 (1996).
\bibitem{vedral97} V. Vedral, M. Plenio, M. Rippin and P. Knight, 
Phys. Rev. Lett. {\bf 78}, 2275 (1997).
\bibitem{wootters97} W. Wootters, Phys.~Rev.~Lett. {\bf 80}, 2245 (1998).
\bibitem{terhal00} B. Terhal and K. Vollbrecht, quant-ph/0005062 (2000).
\bibitem{hughston93} L. Hughston, R. Jozsa and W. Wootters,
Phys. Lett. A {\bf 183}, 14 (1993).
\bibitem{horn85} R. Horn and C. Johnson, {\em Matrix Analysis} 
(Cambridge University Press, 1985).
\bibitem{horodecki97} P. Horodecki, Phys.~Lett.~A {\bf 232}, 333 (1997).
\bibitem{uhlmann} A. Uhlmann, quant-ph/9704017 (1997).
\bibitem{thompson79} R.C. Thompson, Linear Algebra and its Applications 
{\bf 26},65--106 (1979).
\bibitem{dehaene} J. Dehaene, Cheng Yi and B. De Moor, IEEE Trans. 
Automatic Control {\bf 42}, No 11, 1596--1600, 1997.
\bibitem{fletcher} R. Fletcher, {\em Practical Methods of Optimization}, 
Wiley, 1987.
\bibitem{cardoso91} J.-F. Cardoso, Proc. ICASSP'91, Vol. 5, 3109--3112 (1991).
\bibitem{zippel79} R. Zippel, in E. Ng (ed): Proc. of the International 
Symposium on Symbolic and Algebraic Manipulation (EUROSAM '79) Marseille, 
France, June 1979. Lecture Notes in Computer Science {\bf 72}, 216--226 
(Springer, 1979).
\bibitem{horn91} R. Horn and C. Johnson, {\em Topics in Matrix Analysis} 
(Cambridge University Press, 1991).
\end{thebibliography}
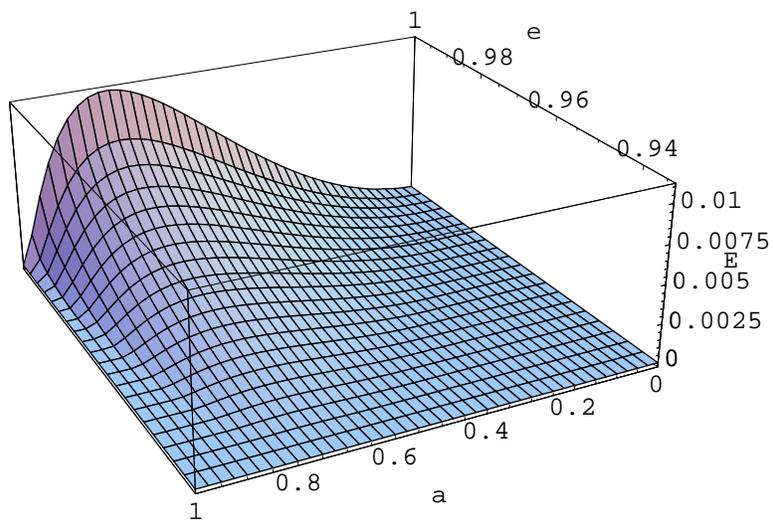
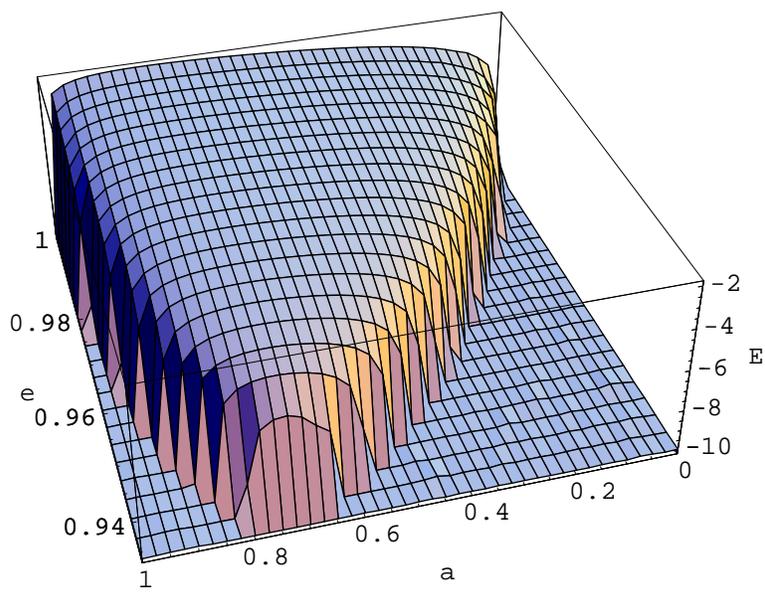
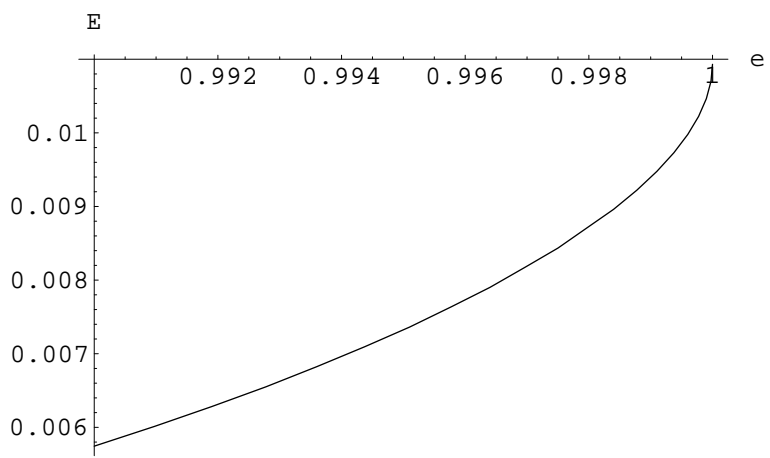
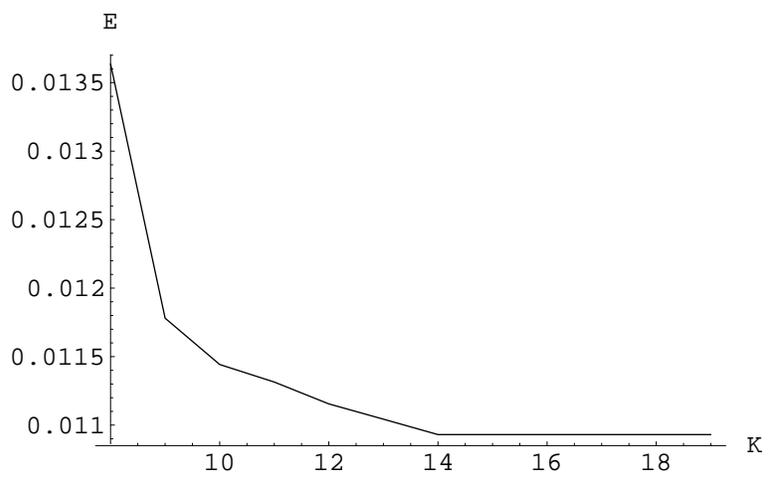
\end{document}